\documentclass[12pt]{iopart}

\begin{document}

\title[Mapping the gravitational wave background]{Mapping the gravitational wave
background}

\author{Neil J. Cornish\footnote[3]{{\tt (cornish@physics.montana.edu)}}}

\address{Department of Physics, Montana State University, Bozeman,
MT 59717, USA}

\begin{abstract}
The gravitational wave sky is expected to have isolated bright sources superimposed on
a diffuse gravitational wave background. The background radiation has two
components: a confusion limited background from unresolved
astrophysical sources; and a cosmological component formed during the birth of the
universe. A map of the gravitational wave background can be made by sweeping
a gravitational wave detector across the sky. The detector output is a complicated
convolution of the sky luminosity distribution, the detector response function and
the scan pattern. Here we study the general de-convolution problem, and show how LIGO
(Laser Interferometric Gravitational Observatory) and
LISA (Laser Interferometer Space Antenna) can be used to detect anisotropies
in the gravitational wave background.
\end{abstract}



\maketitle

\section{Introduction}

Gravitational wave detectors are fairly blunt instruments.
In the low frequency limit,
which corresponding to wavelengths large compared to the effective size
of the detector, a gravitational wave detector's antenna pattern has only monopole,
quadrupole and sextupole components. Despite this limitation, it is in principle possible to
locate a source with arbitrary precision so long as the signal-to-noise ratio
is large and the motion of the detector with respect to the source is sufficiently
fast and varied.

Gravitational wave detectors get their directional information from the amplitude
and frequency modulation that occur as the antenna pattern is swept across the sky. The
frequency modulation is due to the Doppler shift caused by the relative motion of the
source and the detector. For example, orbital motion about the Sun creates a periodic
frequency shift with amplitude $\delta f/f \sim 10^{-4}\sin\theta$, where $\theta$ is
the co-latitude in ecliptic coordinates. The amplitude modulation occurs as the
antenna lobes are swept across the sky. For the LIGO detectors\cite{ligo} this
occurs with a fundamental period of one sidereal day, while for the LISA detector\cite{lisa} the
fundamental period is one sidereal year. When we are lucky enough to have two detectors,
differences in the arrival time can be used to furnish further directional
information. A discussion of the angular resolution of LIGO can be found in the work of
Jaranowski \& Kr\'{o}lak\cite{jk}.
Discussions of the angular resolution of LISA can be found in the work
of Peterseim {\it et al.} \cite{pjds}, Cutler\cite{cc} and Hellings \& Moore\cite{hm}.

In addition to individual bright sources of gravitational waves, there will
also be a variety of unresolved gravitational wave backgrounds. It is likely that
one component of the background will have a cosmological origin - the cosmic
gravitational wave background (CGB). However, the dominant contribution is expected
to be a confusion background formed by the superposition of many weak point sources,
such as white dwarf binaries. The gravitational wave backgrounds will be approximately
stochastic, making them difficult to distinguish from noise in the detectors. One way
around this problem is to cross-correlate the output from two detectors. Since the
noise in each detector is uncorrelated while the signal is correlated, the signal-to-noise
improves as the square root of the observation time. Typically, this type of
correlation has to be done over a finite frequency interval $\Delta f$ that is large
compared to the Doppler modulation $\delta f$, so the frequency
modulation is washed out. Thus, maps of the gravitational wave background
have to be made using amplitude modulation alone.

Here we describe how gravitational wave detectors can be used to create maps of
the gravitational wave background. The pioneering work of Allen and Ottewill\cite{ao} showed
how anisotropies in the gravitational wave background will modulate
the output from the cross-correlated LIGO detectors. A similar calculation was
carried out by Giampieri and Polnorev\cite{gp} in the low frequency limit ($f < 1$ mHz)
for the LISA detector. They showed how the signal would
be modulated by the antenna's orbital motion. Neither group addressed the
inverse problem - that is, how to create a map of the gravitational wave
sky from the modulated output. We address the inverse problem and
find that the ground-based detectors that are
currently under construction have the ability to produce maps with a resolution of
$\ell \sim 20$ at $f=100$ Hz, where $\ell$ is the multipole number. This is comparable to
the maps of the cosmic microwave background (CMB)
produced by the COBE (COsmic Background Explorer) satellite. We find that
a single LISA detector can produce a partial map with a
resolution of $\ell \sim 6$ at $f=10$ mHz, improving to $\ell \sim 25$ at $f=100$ mHz.
More than one LISA detector needs to be flown to produce a complete map. For example,
two LISA detectors with a relative orbital inclination of $\sim 20$ degrees could
produce a complete map with a resolution of $\ell \sim 8$.

A full implementation of the theoretical scheme developed here will be published
elsewhere\cite{cl2}. It involves an end-to-end simulation of the data analysis,
starting with a synthetic gravitational wave sky that produces
a modulated signal from a simulated gravitational wave detector. The
detector model incorporates a basic treatment of the anticipated noise.
Finally, the simulated signal is deconvolved to
produce a sky map that is compared to the original.

\section{The general convolution problem}

Suppose the gravitational wave background has
a luminosity distribution given by ${\cal L}(\theta,\phi,f)$.
Here $(\theta,\phi)$ are coordinates on the sky in the CMB rest frame and $f$ is the
gravitational wave frequency. The response of a gravitational wave detector can be
characterized by a detector response function ${\cal F}(\theta,\phi,f,t)$.
The response varies with time as the detector moves relative to
the sky. The detector output $C(t,f)$ is given by the all-sky integral
\begin{equation}\label{conv}
C(t,f) = \oint_{S^2} {\cal L}(\theta,\phi,f){\cal F}(\theta,\phi,f,t)\,
\sin\theta d\theta d\phi\, .
\end{equation}
Given the time series $C(t,f)$ we need to find the luminosity distribution
${\cal L}(\theta,\phi,f)$.

It is natural to decompose the luminosity distribution and the detector
response function into spherical harmonics in their respective rest frames. Writing
$\vec{e}_\alpha$ for basis vectors in the sky frame and $\vec{e}_{\bar{\alpha}}$
for the detector frame we have
\begin{equation}
\vec{e}_\alpha = \Lambda^{\bar{\beta}}_{\; \alpha}(t)\, \vec{e}_{\bar{\beta}}
\quad \Rightarrow \quad \Lambda_{\alpha\bar{\beta}}(t)
=\vec{e}_\alpha \cdot \vec{e}_{\bar{\beta}} \, ,
\end{equation}
where $\Lambda_{\alpha\bar{\beta}}(t)$ are the components of the coordinate
transformation relating the two frames. In practice the gravitational fields will
be weak and the velocities slow, so the coordinate transformation is well
approximated by a spatial rotation. The coordinates
$(\theta,\phi)$ in the sky frame are related to the coordinates
$(\bar{\theta},\bar{\phi})$ in the detector frame by
\begin{equation}\label{tht}
\cos\bar{\theta} = \widehat{\Omega} \cdot \vec{e}_{\bar{z}} 
= \sin\theta \cos\phi\, \Lambda_{x\bar{z}}+\sin\theta\sin\phi\, \Lambda_{y\bar{z}}
+\cos\theta\, \Lambda_{z\bar{z}} \, ,
\end{equation}
and
\begin{eqnarray}\label{pac}
&& \hspace*{-0.85in} \sin\bar\theta \, e^{i\bar{\phi}}
= \widehat{\Omega}\cdot ( \vec{e}_{\bar{x}} +i
\vec{e}_{\bar{y}}) \nonumber \\ \bs
&&  \hspace*{-0.2in} = \sin\theta\cos\phi (\Lambda_{x\bar{x}}+i\Lambda_{x\bar{y}})
+\sin\theta\sin\phi (\Lambda_{y\bar{x}}+i\Lambda_{y\bar{y}})
+\cos\theta (\Lambda_{z\bar{x}}+i\Lambda_{z\bar{y}})
\end{eqnarray}
We have written the transformations in this form since these are the combinations that appear
in the  spherical harmonics:
\begin{equation}\label{sph}
Y_{\ell m}(\theta,\phi) = \sqrt{\frac{2\ell + 1}{4\pi}\frac{(\ell -m)!}{(\ell
+m)!}}\left(-\sin\theta \, e^{i\phi}\right)^m \frac{d^m}{d(\cos\theta)^m}
P_{\ell}(\cos\theta) \, .
\end{equation}
Decomposing ${\cal L}(\theta,\phi,f)$ and ${\cal F}(\theta,\phi,f,t)$ in
the sky frame we have
\begin{equation}\label{decom}
\fl {\cal L}(\theta,\phi,f) = \sum_{\ell m} p_{\ell m}(f) Y_{\ell m}(\theta,\phi) \, , \quad
{\rm and } \quad 
{\cal F}(\theta,\phi,f,t) = \sum_{\ell m} a_{\ell m}(f,t) Y_{\ell m}(\theta,\phi)\, ,
\end{equation}
while in the detector's restframe we have
\begin{equation}
{\cal F}(\bar{\theta},\bar{\phi},f) = \sum_{\ell m} \bar{a}_{\ell m}(f) Y_{\ell m}
(\bar{\theta},\bar{\phi})\, .
\end{equation}
The multipole moments in the two frames are related by
\begin{equation}
a_{\ell m}(f,t) = \sum_{n=-\ell}^{\ell} \bar{a}_{\ell n}(f) \lambda_{\ell m n}(t) \, ,
\end{equation}
where
\begin{equation}
\lambda_{\ell\, m n}(t) =  \oint_{S^2} 
Y^*_{\ell m}(\theta, \phi) Y_{\ell n}(\bar\theta,\bar\phi)\, d\Omega \, .
\end{equation}
Using equations (\ref{tht}), (\ref{pac}) and (\ref{sph}) it is a simple matter
to calculate the transformation coefficients $\lambda_{\ell m n}(t)$. From the
properties of the spherical harmonics it follows that
\begin{equation}
\lambda^*_{\ell\, m n} = (-1)^{m+n} \lambda_{\ell\, -m\, -n}\, .
\end{equation}
Thus, there we only need to calculate $(2\ell+1)(\ell+1)$ components at 
order $\ell$. Out to order $\ell=1$ we have
\begin{eqnarray}
&& \lambda_{000}(t) = 1  \nonumber \\
&& \lambda_{1-10}(t) = \frac{1}{\sqrt{2}}\left(\Lambda_{xz}(t)+i\Lambda_{yz}(t)\right)
 \nonumber \\
&& \lambda_{1-11}(t) = \frac{1}{2}\left(-\Lambda_{xx}(t)+\Lambda_{yy}(t)
-i \Lambda_{xy}(t)-i \Lambda_{yx}(t) \right) \nonumber \\
&&  \lambda_{100}(t) = \Lambda_{zz}(t)  \nonumber \\
&&  \lambda_{110}(t) = \frac{1}{\sqrt{2}}\left(-\Lambda_{xz}(t)
+i\Lambda_{yz}(t)\right) \nonumber \\
&&  \lambda_{101}(t) = -\frac{1}{\sqrt{2}}\left(\Lambda_{zx}(t)+i\Lambda_{zy}(t)\right)
 \nonumber \\
&&  \lambda_{111}(t) = \frac{1}{2}\left(\Lambda_{xx}(t)+\Lambda_{yy}(t)
+i \Lambda_{xy}(t)-i \Lambda_{yx}(t) \right) \, .
\end{eqnarray}
Higher orders in $\ell$ involve higher powers of $\Lambda_{ij}(t)$. For example,
\begin{equation}
\lambda_{210}(t) = \frac{\sqrt{6}}{2}\Lambda_{zz}(t)\left(-\Lambda_{xz}(t)
+i\Lambda_{yz}(t)\right) \, .
\end{equation}
Using the decompositions defined in (\ref{decom}), we can express the detector
output in terms of the multipole moments $p_{\ell m}(f)$ and
$a_{\ell m}(t)$:
\begin{equation}
C(t,f) = \sum_{\ell, m} (-1)^m\, p_{\ell m}(f)\,  a_{\ell\, -m}(t,f) \, .
\end{equation}
In most instances the detector scan pattern will be periodic in time, so
it is natural to Fourier transform all time dependent quantities:
\begin{equation}
A_k = \frac{1}{2\pi} \int_0^{2\pi} d\alpha \, e^{-ik\alpha} A(\alpha) \, ,
\end{equation}
where $\alpha = 2\pi t/T_0$, and $T_0$ is the period of the detector sweep.
We can write
\begin{equation}\label{genconv}
C_k(f) = \sum_{\ell, m} \, p_{\ell m}(f)\, \gamma_{k\ell m}(f) \,
\end{equation}
where
\begin{equation}
\gamma_{k\ell m}(f) = (-1)^m a_{k \ell\, -m}(f) =  (-1)^m \sum_{n=-\ell}^{\ell}
\bar{a}_{\ell n}(f)\, \lambda_{k\ell\, -m n} \, .
\end{equation}
It is a simple matter to calculate the $\gamma_{k\ell m}$ if we know the detector
response function ${\cal F}$ and the scan pattern $\Lambda_{ij}(t)$. The
de-convolution problem comes down to solving the linear system of equations
\begin{equation}
C_k(f) = p_i(f)\, \gamma_{ki}(f) \quad {\rm where} \quad i=\ell^2+\ell+m+1 \, .
\end{equation}
The $\gamma_{ki}(f)$ are known, the $C_k(f)$ are measured experimentally, and the
$p_i(f)$ are what we want to find. Typically the system of equations will be
both under and over constrained, but a best fit solution can be found by way of a
singular value decomposition.

\section{Map making with LIGO}

We begin by studying how LIGO can be used to map the gravitational wave sky as a
prelude to studying the space-based LISA detector. Allen and Ottewill\cite{ao} showed how the
cross-correlated response of the Hanford and Livingston LIGO detectors is modulated
as the rotation of the Earth sweeps
the antenna pattern across the sky. However, they stopped short of solving
the deconvolution problem to make a map of the gravitational wave sky.

In a sky-fixed reference frame the response of a single LIGO detector to
a plane gravitational wave ${\bf h}(t,{\bf x},f)$ with frequency $f$, propagating
in the $\widehat{\Omega}$ direction is given by
\begin{equation}\label{basic}
s(t) = {\bf D}(t):{\bf h}(t,{\bf x}) \, ,
\end{equation}
where
\begin{equation}
{\bf D}(t) = \frac{1}{2}\left({\bf u}(t)\otimes{\bf u}(t)
-{\bf v}(t)\otimes{\bf v}(t)\right)
\end{equation}
is the detector tensor and ${\bf u}(t)$ and ${\bf v}(t)$
are unit vectors in the direction of the interferometer arms. The response to
a general gravitational wave ${\bf h}(t,{\bf x})$ can be derived from (\ref{basic})
by first decomposing the wave into a collection of plane waves:
\begin{equation}
    {\bf h}(t,{\bf x}) = \sum_{A=+,\times}
    \int_{-\infty}^{\infty} df \int_{S^2} d\widehat{\Omega} \;
    \tilde{h}_A(f,\widehat{\Omega})e^{-2\pi i ft}e^{2\pi i f
    \widehat{\Omega}\cdot{\bf x}/c}
    {\bf e}^{A}(\widehat{\Omega}) \, .
\end{equation}
Here ${\bf e}^{+}$ and ${\bf e}^{\times}$ are a set of polarization tensors.
The signal we wish to analyze is formed by cross-correlating the output
of the two LIGO detectors over an integration time $T$, centered at time $t$:
\begin{equation}
S(t) = \int_{t-T/2}^{t+T/2} dt' \int_{t-T/2}^{t+T/2} dt'' s_1(t') s_2(t'') Q(t'-t'') \, .
\end{equation}
Here $s_1(t)$ and $s_2(t)$ are the outputs of the two detectors and $Q$ is some filter
function. A stochastic background of gravitational waves can be characterized by
the statistical character of the Fourier amplitudes $\tilde{h}_A(f,\widehat{\Omega})$.
Following Allen and Ottewill we assume that the background can be approximated as a
stationary, Gaussian random distribution characterized by the expectation values
\begin{eqnarray}
&& \langle \tilde{h}_A(f,\widehat{\Omega})\rangle =  0 \nonumber \\
&& \langle
    \tilde{h}^*_A(f,\widehat{\Omega})
    \tilde{h}_{A'}(f',\widehat{\Omega}')\rangle
    =  \frac{1}{2}\delta(f-f')
    \frac{\delta^2(\widehat{\Omega},\widehat{\Omega}')}
    {4\pi}\delta_{AA'} \, S_h(f)P(\widehat{\Omega}) \, ,
\end{eqnarray}
where $S_h(f)$ is the power spectral density and $P(\widehat{\Omega})$ describes the
angular distribution. From these relations it follows that the expectation value of
$S(t)$ is given by
\begin{equation}
C(t) = \langle S(t) \rangle = \frac{T}{5} \int^{\infty}_{-\infty} df
    \, S_h(f) \gamma(f,t) \widetilde{Q}(f) \, ,
\end{equation}
where
\begin{equation}
\gamma(f,t) = \frac{5}{8\pi} \int
d\widehat{\Omega} \, P(\widehat{\Omega}) \sum_{A=+,\times}
F_1^{A}(\widehat{\Omega},t)F_2^{A}(\widehat{\Omega},t)
    e^{-2\pi i f \widehat{\Omega}\cdot\Delta{\bf x}(t)/c} \, .
\end{equation}
Here $\Delta{\bf x}(t)$ is a vector connecting the Hanford and Livingston sites and
\begin{equation}
F_i^{A}(\widehat{\Omega},t) = {\bf D}_i(t): {\bf e}^{A} \, .
\end{equation}

Using $S_h(f)=S_h(-f)$, $\widetilde{Q}(f)=\widetilde{Q}^*(-f)$ and $\gamma(f,t)=\gamma^*(-f,t)$,
we can write the signal as
\begin{equation}
C(t)=  \frac{2T}{5}\int^{\infty}_{0} df\, S_h(f)\left(\gamma_R(f,t)\widetilde{Q}_R(f)-
\gamma_I(f,t)\widetilde{Q}_I(f) \right) , ,
\end{equation}
where $\gamma_R(f,t)$ denotes the real part of $\gamma(f,t)$ {\it etc.}.
In contrast to Allen and Ottewill, who considered the broad-band response, we are
interested in making narrow band measurements over a small frequency interval of
width $\Delta f$ centered at frequency $f$. This allows us to make maps of the
gravitational wave sky at particular frequencies. To this end we use the
top-hat filter
\begin{equation}
\widetilde{Q}(f') =
\left \{\begin{array}{ll}
	q_R+ i\, q_I & \quad f-\frac{\Delta f}{2} \leq f' \leq f+\frac{\Delta f}{2}\\ \bs
        q_R- i\, q_I & \quad -f-\frac{\Delta f}{2} \leq f' \leq -f+\frac{\Delta f}{2}\\ \bs
        0 &  \quad {\rm otherwise} 
	\end{array} \right.
\end{equation}
so that
\begin{equation}
Q(t) = 2 \frac{\sin(\pi t \Delta f)}{\pi t}\left( q_R \cos(2\pi f t)+q_I\sin(2\pi f t)\right).
\end{equation}
In the time domain the filter decays rapidly for times $t>1/\Delta f$. In
choosing $\Delta f$ we have
to ensure that $1/\Delta f \ll T$, otherwise the angular variation will be smeared out. 
So long as $\Delta f$ is small compared to the range over which $S_h(f)$ and
$\gamma(f,t)$ vary, the narrow band response is given by
\begin{equation}
C(t,f) = \frac{2T\Delta f}{5} \, S_h(f)\, \left(q_R \gamma_R(f,t)-q_I \gamma_I(f,t)\right) \, .
\end{equation}
This equation is of the form
\begin{equation}
C(t,f) = \oint_{S^2} {\cal L}(\theta,\phi,f){\cal F}(\theta,\phi,f,t)\,
\sin\theta d\theta d\phi \, ,
\end{equation}
where the luminosity equals
\begin{equation}
{\cal L}(\theta,\phi,f) =  S_h(f) P(\widehat{\Omega}) \, ,
\end{equation}
and the detector response function is given by
\begin{eqnarray}
{\cal F}(\theta,\phi,f,t) &=& \frac{T\Delta f q_R}{4\pi} \sum_{A=+,\times}
F_1^{A}(\widehat{\Omega},t)F_2^{A}(\widehat{\Omega},t)
    \cos\left(2\pi f \widehat{\Omega}\cdot\Delta{\bf x}(t)/c\right) \nonumber \\ \bs
&+& \frac{T\Delta f q_I}{4\pi} \sum_{A=+,\times}
F_1^{A}(\widehat{\Omega},t)F_2^{A}(\widehat{\Omega},t)
    \sin\left(2\pi f \widehat{\Omega}\cdot\Delta{\bf x}(t)/c\right)
\end{eqnarray}
The choice of integration period $T$ results from a trade-off between maximizing the
signal ($T$ large) and maximizing the angular resolution ($T$ small).
Setting a maximum angular resolution of $k_{\rm max}$ fixes the integration period
to be (Nyquist's theorem)
\begin{equation}
T = \frac{T_0}{2 k_{\rm max}} \, .
\end{equation}
It turns out that there is little to be gained by setting $k_{\rm max}$ much larger
than 50, so we will take $k_{\rm max}=50$ and set $T=861.64$ s. Similar considerations
apply to the choice of $\Delta f$, which for LIGO we will take to be in the range
$\Delta f = 1 \rightarrow 10$ Hz.
A plot of the antenna pattern for the cross-correlated pair of LIGO detectors is shown
in Figure 1 for a range of frequencies. As expected, the antenna pattern has more structure at
high frequencies.

\begin{figure}[ht]
\vspace*{3.3in}
\includegraphics{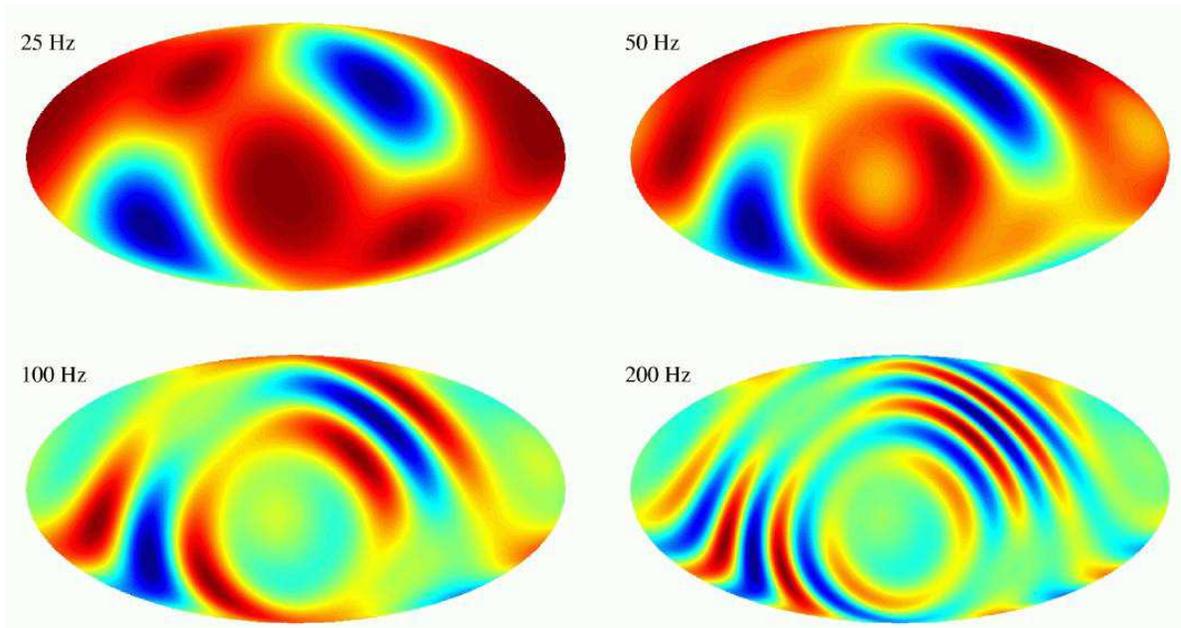}
\caption{The antenna pattern for the cross-correlated LIGO detectors in the Earth-fixed frame,
${\cal F}(\bar{\theta},\bar{\phi},f)$, with $g_I=0$ and $f=25,50,100$ and 200 Hz. }
\end{figure} 

Now that we have the LIGO signal expressed in terms of the general convolution problem of
the last section, it is a simple matter to
carry through the analysis. We begin by setting $\alpha=2\pi t /T_d$ where $T_d=86164$ s is
one sidereal day. Choosing Sky-fixed and Earth-fixed coordinate systems with their $z$ axes
aligned with the Earth's rotation axis, we find the two frames are related:
\begin{eqnarray}
\vec{e}_{\bar x} &=&  \cos\alpha\, \vec{e}_x + \sin\alpha\, \vec{e}_y \nonumber \\
\vec{e}_{\bar y} &=&  -\sin\alpha\, \vec{e}_x + \cos\alpha\, \vec{e}_y \nonumber \\
\vec{e}_{\bar z} &=& \vec{e}_z \, ,
\end{eqnarray}
where the bar denotes the Earth-fixed frame. From these relations it follows that
\begin{equation}
\lambda_{k\ell m n} = \delta_{mn}\, \delta_{km} ,
\end{equation}
and
\begin{equation}
\gamma_{klm}(f) = \delta_{km} (-1)^m\bar{a}_{\ell \, -m}(f) =  \delta_{km} \gamma_{\ell m}(f) \, .
\end{equation}
Here $\gamma_{\ell m}(f)$ correspond to the quantities calculated by Allen and Ottewill. However,
they evaluated $\gamma_{\ell m}(f)$ in an Earth-fixed frame with the $z$ axis
parallel to the vector connecting the two detectors, $\Delta {\bf x}$. A static rotation
has to be applied to their results before we can use them. Once this is done we can
attack the inversion problem
\begin{equation}\label{ligo}
C_k(f) = \sum_{\ell m} p_{\ell m}(f) \gamma_{k\ell m}(f) 
=   \sum_{\ell = \vert k\vert}^\infty
p_{\ell k}(f) \, \gamma_{\ell k}(f)  \, ,
\end{equation}
where the final line was obtained from the identity
\begin{equation}
\sum_{\ell m} = \sum_{\ell = 0}^{\infty}\sum_{m = -\ell}^{\ell} = \sum_{m=-\infty}^{\infty}
\sum_{\ell=\vert m\vert}^{\infty} \, .
\end{equation}

\begin{figure}[ht]
\vspace*{4.8in}
\includegraphics{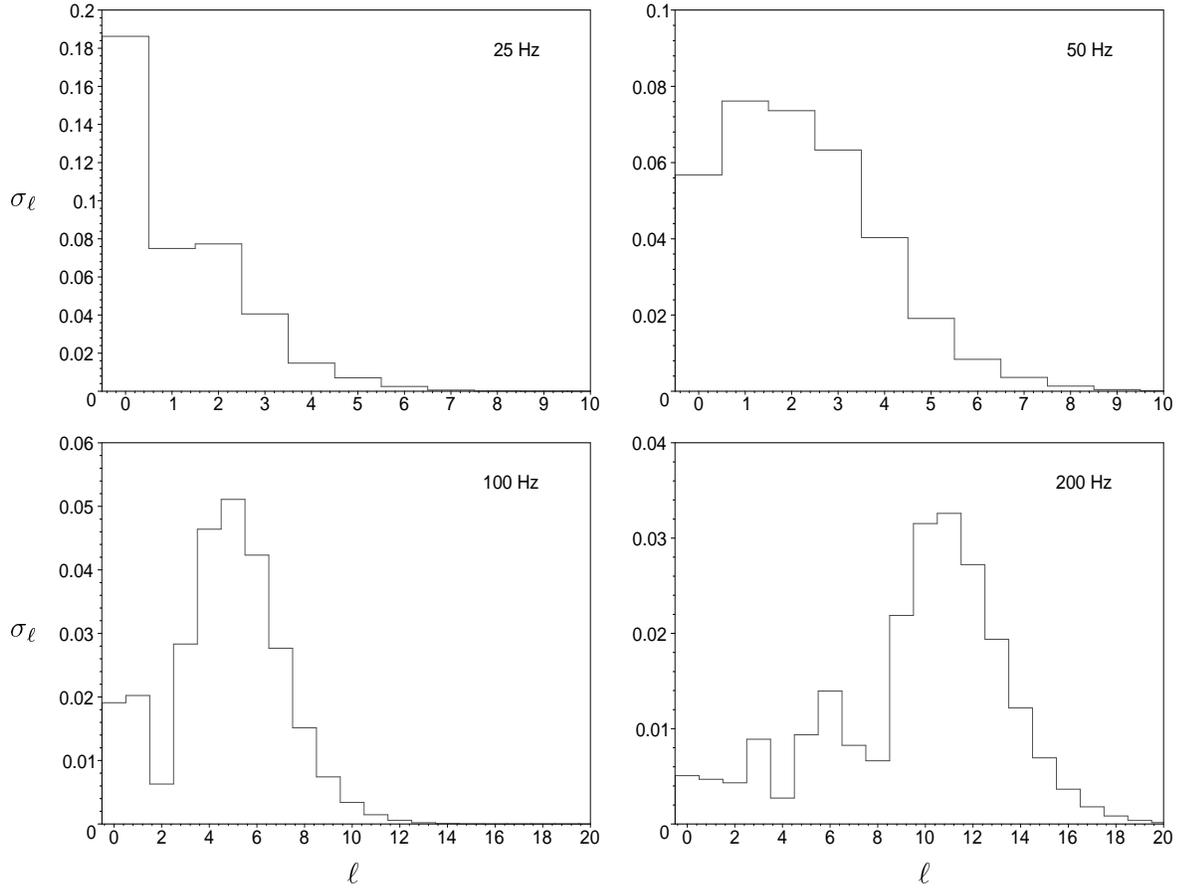}
\caption{The root-mean-square amplitudes of $\gamma_{\ell m}(f)$ for a range of
frequencies.}
\end{figure}

At first sight it appears that we have no hope of solving the inversion problem since
it involves $2(2k_{\rm max}+1)$ equations\footnote[2]{The equation count follows from
$-k_{\rm max} \leq k \leq k_{\rm max}$ and the freedom to alternately set $q_R$
and $q_I$ equal to zero.} and an infinite number of unknowns. However,
the $\gamma_{\ell m}(f)$ decay rapidly for large $\ell$, effectively
reducing the number of unknowns. The root-mean-square amplitude of the $\gamma_{\ell m}(f)$'s,
\begin{equation}
\sigma_\ell(f) =\sqrt{C_{\ell}(f)} \, ,
\end{equation}
provides a good estimate of how each order in $\ell$ contributes to the sum in (\ref{ligo}).
Here $C_\ell(f)$ is the usual angular power spectrum:
\begin{equation}
C_\ell(f) = \frac{1}{2\ell +1}
\sum_{m =-\ell}^{\ell} \vert \gamma_{\ell m}(f) \vert^2\, .
\end{equation}
Plots of $\sigma_\ell(f)$ for $f=25,50,100$ and 200 Hz are shown in Figure 2. In each case,
the low multipoles dominate the response. Thus, the sum (\ref{ligo}) can be approximated by
\begin{equation}\label{ligotrunc}
C_k(f) = \sum_{\ell =\vert k\vert}^{\ell_{\rm max}}
p_{\ell k}(f) \, \gamma_{\ell k}(f)  \, ,
\end{equation}
where $\ell_{\rm max} \approx 5$ for $f=25$ Hz, rising to $\ell_{\rm max} \approx 18$ for
$f=200$ Hz. The number of equations is restricted to those with
$\vert k \vert \leq \ell_{\rm max}$. If we write $C_k = q_R A_k + q_I B_k$ and use two
sets of filters, one with $q_R=1$, $q_I=0$, and the other with $q_R=0$ and $q_I=1$, then
(\ref{ligotrunc}) can be written as a set of $4\ell_{\rm max}$ equations for
$(\ell_{\rm max} +1)^2$ unknowns. As an example, consider the inversion problem at 25 Hz,
where the system of equations is (for $k\geq 0$):
\begin{eqnarray}
A_0 &= p_{00}\gamma_{00} + p_{20}\gamma_{20}+p_{40}\gamma_{40}
\quad & B_0 = p_{10}\gamma_{10} + p_{30}\gamma_{30}+p_{50}\gamma_{50} \nonumber \\
A_1 &= p_{21}\gamma_{21} + p_{41}\gamma_{41}
& B_1 = p_{11}\gamma_{11} + p_{31}\gamma_{31}+p_{51}\gamma_{51} \nonumber \\
A_2 &= p_{22}\gamma_{22} + p_{42}\gamma_{42}
&B_2 = p_{32}\gamma_{32}+p_{52}\gamma_{52} \nonumber \\
A_3 &= p_{43}\gamma_{43}
&B_3 = p_{33}\gamma_{33}+p_{53}\gamma_{53} \nonumber \\
A_4 &= p_{44}\gamma_{44}
&B_4 = p_{54}\gamma_{54} \nonumber \\
&& B_5 = p_{55}\gamma_{55} \, .
\end{eqnarray}
The above system of 20 equations involves 36 $p_{\ell m}$'s, making it impossible to
solve for each moment individually. The way around this impasse is to build
another interferometer at a different location and to cross-correlate its output with
the two LIGO detectors, in effect tripling the number of equations. Currently there
are several interferometers being built in addition to the LIGO pair, including the
Italian-French VIRGO detector near Pisa, the Anglo-German GEO detector near Hanover,
and the Japanese TAMA detector near Tokyo. Cross-correlating the signals from
all of these detectors will give roughly $40\ell_{\rm max}$ equations
for the $(\ell_{\rm max} +1)^2$ unknowns, making it possible to create maps of the
gravitational wave background with a resolution as high as $\ell_{\rm max} \sim 38$.
Indeed, the limitation on the resolution will not come from the system being
under determined, but from the lack of antenna sensitivity at high $\ell$.

\section{Map making with LISA}

The procedure for making a map of the gravitational wave sky with the LISA detector is
very similar to the procedure we used for the LIGO detectors. Currently there are
no plans to fly two sets of LISA spacecraft, so we will probably have to make do with
the self-correlated signal from a single detector. At first sight it would appear that
correlated noise in the self-correlated signal will prevent us from being able to distinguishing
detector noise from a stochastic gravitational wave signal. However, this turns out not
to be the case. Firstly, it is possible to discriminate between detector noise and
a stochastic signal by using the Sagnac signal that is formed by comparing the phase of
a signal that is sent clockwise around the LISA triangle with a signal that is
sent counter-clockwise\cite{tae,hb}. The Sagnac signal is very insensitive to gravitational waves,
making it the perfect tool for monitoring instrument noise. Secondly, the signal
will vary periodically as the detector sweeps across the sky whereas the noise
will not. Thus, by making several complete sweeps, it is possible to build up the
signal to noise ratio in the measurements of $C_k$ for all $k\neq 0$. We
study the map making capabilities of a single LISA interferometer and a
pair of optimally cross-correlated LISA interferometers. The optimal cross-correlation
is achieved by placing six LISA spacecraft in a circle, with two sets of
three spacecraft forming independent interferometers rotated by 90 degrees in
the plane of the circle\cite{me}.

Proceeding as we did in the last section, the cross-correlated signal is given by
\begin{equation}
C(t,f) = \oint_{S^2} {\cal L}(\theta,\phi,f){\cal F}(\theta,\phi,f,t)\,
\sin\theta d\theta d\phi \, ,
\end{equation}
where the luminosity is again given by
\begin{equation}
{\cal L}(\theta,\phi,f) =  S_h(f) P(\widehat{\Omega}) \, ,
\end{equation}
and the detector response function has the form
\begin{eqnarray}
{\cal F}(\theta,\phi,f,t) &=& \frac{T\Delta f q_R}{4\pi} \Re \left[ \sum_{A=+,\times}
F_1^{A}(\widehat{\Omega},f,t){F_2^{A}}^{*}(\widehat{\Omega},f,t)
    e^{2\pi i f \widehat{\Omega}\cdot\Delta{\bf x}(t)/c}\right] \nonumber \\ \bs
 &-& \frac{T\Delta f q_I}{4\pi} \Im \left[ \sum_{A=+,\times}
F_1^{A}(\widehat{\Omega},f,t){F_2^{A}}^{*}(\widehat{\Omega},f,t)
    e^{2\pi i f \widehat{\Omega}\cdot\Delta{\bf x}(t)/c} \right].
\end{eqnarray}
While the basic expressions are similar for LISA and LIGO, the LISA
antenna patterns are considerably more complicated:
\begin{equation}
F_i^A(\widehat{\Omega},f,t)={\bf D}_i(\widehat{\Omega},f,t):{\bf e}^{A}(\widehat{\Omega})
\end{equation}
where
\begin{equation}\label{dresp}
\fl {\bf D}(\widehat\Omega,f,t) = \frac{1}{2}\left( ({\bf u}(t)\otimes{\bf u}(t))
        \, {\cal T}({\bf u}\cdot\widehat{\Omega},f) -
    ({\bf v}(t)\otimes{\bf v}(t))\, {\cal T}({\bf
    v}\cdot\widehat{\Omega},f) \right)  \, ,
\end{equation}
is the detector tensor and
\begin{eqnarray}
    {\cal T}({\bf a}\cdot\widehat{\Omega},f)&=&\frac{1}{2}\left[
    {\rm sinc}\left( \frac{f}{2f_* }(1-{\bf
    a}\cdot\widehat{\Omega})\right)\exp\left(-i\frac{f}{2f_*}(3 +{\bf
    a}\cdot\widehat{\Omega})\right) \right. \nonumber \\
    &+& \left. {\rm sinc}\left(\frac{f}{2f_* }(1+{\bf
    a}\cdot\widehat{\Omega})\right) \exp\left(-i\frac{f}{2f_*}(1+{\bf
    a}\cdot\widehat{\Omega})\right)\right] \, ,
\end{eqnarray}
is the transfer function\cite{cl}. The transfer frequency, $f_*=c/(2\pi L)$, corresponds to a
wave that just fits inside the interferometer. The LISA interferometer will have arms of length
$L =5\times 10^6$ km.
The antenna pattern sweeps over the sky as the LISA constellation orbits about the Sun with
period $T_\odot=1$ sidereal year. Setting $k_{\rm max}=50$ fixes the integration period
to be $T=315581$ s (roughly three and a half days). 

\begin{figure}[ht]
\vspace*{3.3in}
\includegraphics{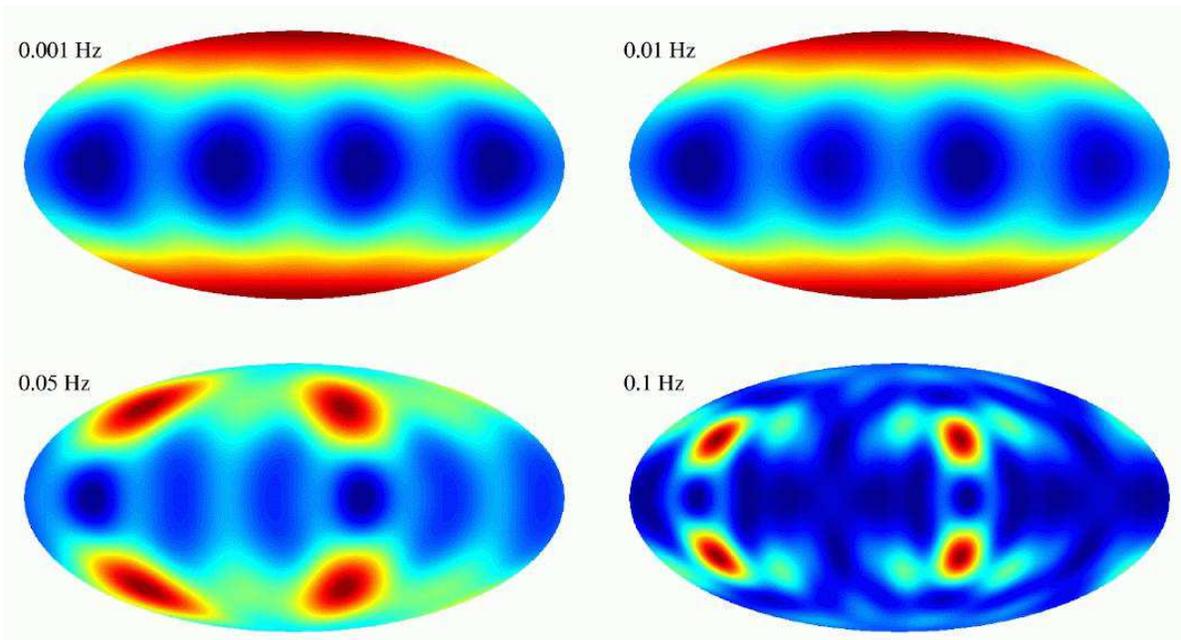}
\caption{The antenna pattern for the self-correlated LISA detector in the detector
rest frame for a range of frequencies.}
\end{figure} 

Choosing Sky-fixed and LISA-fixed coordinate systems with their $z$ axes normal to
the ecliptic gives two frames of reference related by
\begin{eqnarray}
\vec{e}_{\bar x} &=&  \frac{1}{\sqrt{8}}(2-\cos^2\alpha-\cos\alpha\sin\alpha)\vec{e}_x
+ \frac{1}{\sqrt{8}}(1+\cos^2\alpha-\cos\alpha\sin\alpha)\vec{e}_y \nonumber \\
&& -\frac{\sqrt{3}}{\sqrt{8}}(\cos\alpha+\sin\alpha)\vec{e}_z \nonumber \\ \bs
\vec{e}_{\bar y} &=&  \frac{1}{\sqrt{8}}(-2+\cos^2\alpha-\cos\alpha\sin\alpha)\vec{e}_x
+ \frac{1}{\sqrt{8}}(1+\cos^2\alpha+\cos\alpha\sin\alpha)\vec{e}_y \nonumber \\
&& +\frac{\sqrt{3}}{\sqrt{8}}(\cos\alpha-\sin\alpha)\vec{e}_z \nonumber \\ \bs
\vec{e}_{\bar z} &=& \frac{\sqrt{3}}{2}\cos\alpha \vec{e}_x 
+ \frac{\sqrt{3}}{2}\sin\alpha\vec{e}_y+\frac{1}{2} \vec{e}_z \, ,
\end{eqnarray}
where the bar denotes the LISA-fixed frame. The complicated scan pattern means that we
have to calculate each rotation coefficient $\lambda_{k\ell mn}$ individually. Since
the coordinate transformation $\Lambda_{ij}(\alpha)$ is second order in $e^{i \alpha}$,
the index $k$ will run from $-2\ell$ to $2\ell$. To order $\ell=2$ the rotation coefficients are:
\begin{equation}
\fl \begin{array}{llll}
&\lambda_{0000}=1  \quad\quad 
&\lambda_{01-1-1}=\frac{3\sqrt{2}}{8}\left(1+i\right)\quad\quad  
&\lambda_{11-10}=\frac{\sqrt{6}}{4}  \\ \bs
&\lambda_{21-11}= \frac{\sqrt{2}}{8}\left(1-i\right) \quad\quad 
&\lambda_{-110-1}=-\frac{\sqrt{3}}{4}\left(1+i\right)\quad\quad 
&\lambda_{0100}=\frac{1}{2}  \\ \bs
&\lambda_{1101}= \frac{\sqrt{3}}{4}\left(1-i\right)   \quad\quad
&\lambda_{-211-1}= \frac{\sqrt{2}}{8}\left(1+i\right)  \quad\quad 
&\lambda_{-1110}= -\frac{\sqrt{6}}{4}  \\ \bs
&\lambda_{0111}=  \frac{3\sqrt{2}}{8}\left(1-i\right) \quad\quad
&\lambda_{02-2-2}= \frac{9i}{16}  \quad\quad 
&\lambda_{12-2-1}= \frac{3\sqrt{6}}{16}\left(1+i\right)  \\ \bs
&\lambda_{22-20}= \frac{3\sqrt{150}}{80}   \quad\quad
&\lambda_{32-21}= \frac{\sqrt{6}}{16}\left(1-i\right)  \quad\quad 
&\lambda_{42-22}= -\frac{i}{16}  \\ \bs
&\lambda_{-12-1-2}= -\frac{3\sqrt{3}i}{8}   \quad\quad
&\lambda_{12-10}= \frac{3\sqrt{2}}{8}  \quad\quad 
&\lambda_{22-11}= \frac{\sqrt{2}}{4}\left(1-i\right)  \\ \bs
&\lambda_{32-12}= -\frac{\sqrt{3}i}{8}    \quad\quad
&\lambda_{-220-2}= \frac{3\sqrt{150}i}{80}   \quad\quad 
&\lambda_{-120-1}= -\frac{3}{8}\left(1+i\right) \\ \bs
&\lambda_{0200}=  -\frac{1}{8}  \quad\quad
&\lambda_{1201}= \frac{3}{8}\left(1-i\right)   \quad\quad 
&\lambda_{2202}=  -\frac{3\sqrt{150}i}{80}  \\ \bs
&\lambda_{-321-2}= -\frac{\sqrt{3}i}{8}    \quad\quad
&\lambda_{-221-1}= \frac{\sqrt{2}}{4}\left(1+i\right)  \quad\quad 
&\lambda_{-1210}= -\frac{3\sqrt{2}}{8}  \\ \bs
&\lambda_{1212}= -\frac{3\sqrt{3}i}{8}   \quad\quad
&\lambda_{-422-2}= \frac{i}{16}  \quad\quad 
&\lambda_{-322-1}= - \frac{\sqrt{6}}{16}\left(1+i\right)  \\ \bs
&\lambda_{-2220}= \frac{3\sqrt{150}}{80}  \quad\quad
&\lambda_{-1221}=  -\frac{\sqrt{6}}{16}\left(1-i\right) \quad\quad 
&\lambda_{0222}= -\frac{9i}{16}  
\end{array}
\end{equation}

Then antenna patterns for a self-correlated LISA detector and an optimally cross-correlated
pair of LISA interferometers are shown in Figures 3, 4 and 5 for a range of frequencies in
the rest frame of the detectors. These patterns can be turned into $\bar{a}_{\ell m}$'s
using the {\it HEALPIX} software package\cite{hp}. In combination with the analytically
derived rotation coefficients, $\lambda_{k \ell m n}$, the $\bar{a}_{\ell m}$'s yield the
$\gamma_{k\ell m}$'s that appear in the deconvolution problem (\ref{genconv}).
In the low frequency limit, $f \ll f_*$, it is possible to derive the antenna harmonics
$\bar{a}_{\ell m}$ analytically. For the self-correlated LISA detector the non-vanishing
harmonics at zero frequency are:
\begin{equation}
\begin{array}{l}
\bar{a}_{00}= 1 \\ \bs
\bar{a}_{20}= \frac{2\sqrt{5}}{7} \\ \bs
\bar{a}_{4-4}= \frac{\sqrt{70}}{84}+i\frac{\sqrt{210}}{252} \\ \bs
\bar{a}_{4 0}= \frac{1}{42} \\ \bs
\bar{a}_{44}= \frac{\sqrt{70}}{84}-i\frac{\sqrt{210}}{252} 
\end{array}
\end{equation}
while for the optimally cross-correlated pair of LISA detectors
\begin{equation}
\begin{array}{l}
\bar{a}_{00}= -1 \\ \bs
\bar{a}_{20}= -\frac{2\sqrt{5}}{7} \\ \bs
\bar{a}_{4-4}= -\frac{\sqrt{70}}{84}\\ \bs
\bar{a}_{4 0}= -\frac{1}{42} \\ \bs
\bar{a}_{44}= -\frac{\sqrt{70}}{84} \, . \\
\end{array}
\end{equation}

\begin{figure}[ht]
\vspace*{3.3in}
\includegraphics{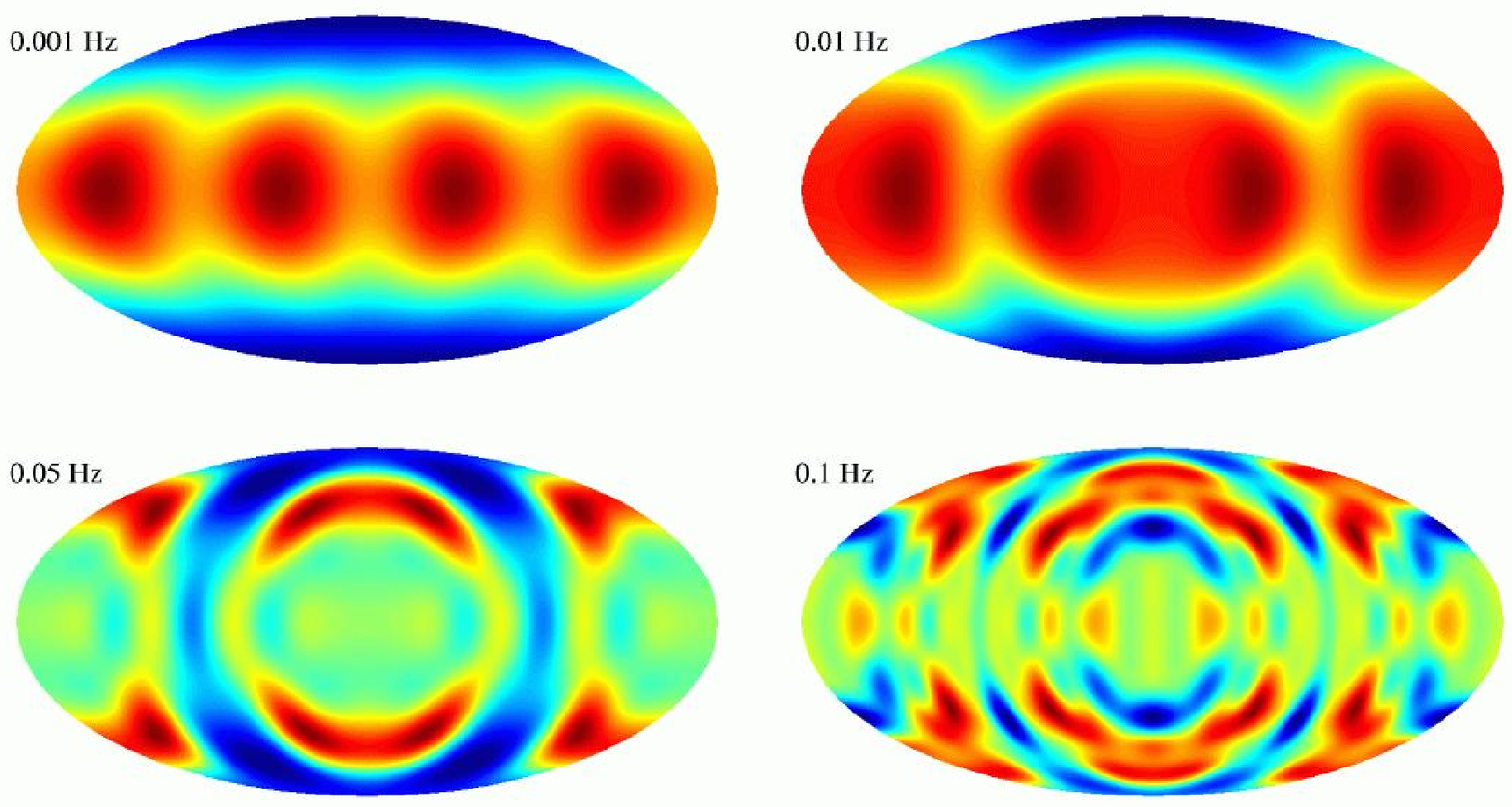}
\caption{The antenna pattern for two optimally cross-correlated LISA detectors in the detector
rest frame for a range of frequencies. Here $q_R=1$ and $q_I=0$.}
\end{figure} 

\begin{figure}[ht]
\vspace*{3.3in}
\includegraphics{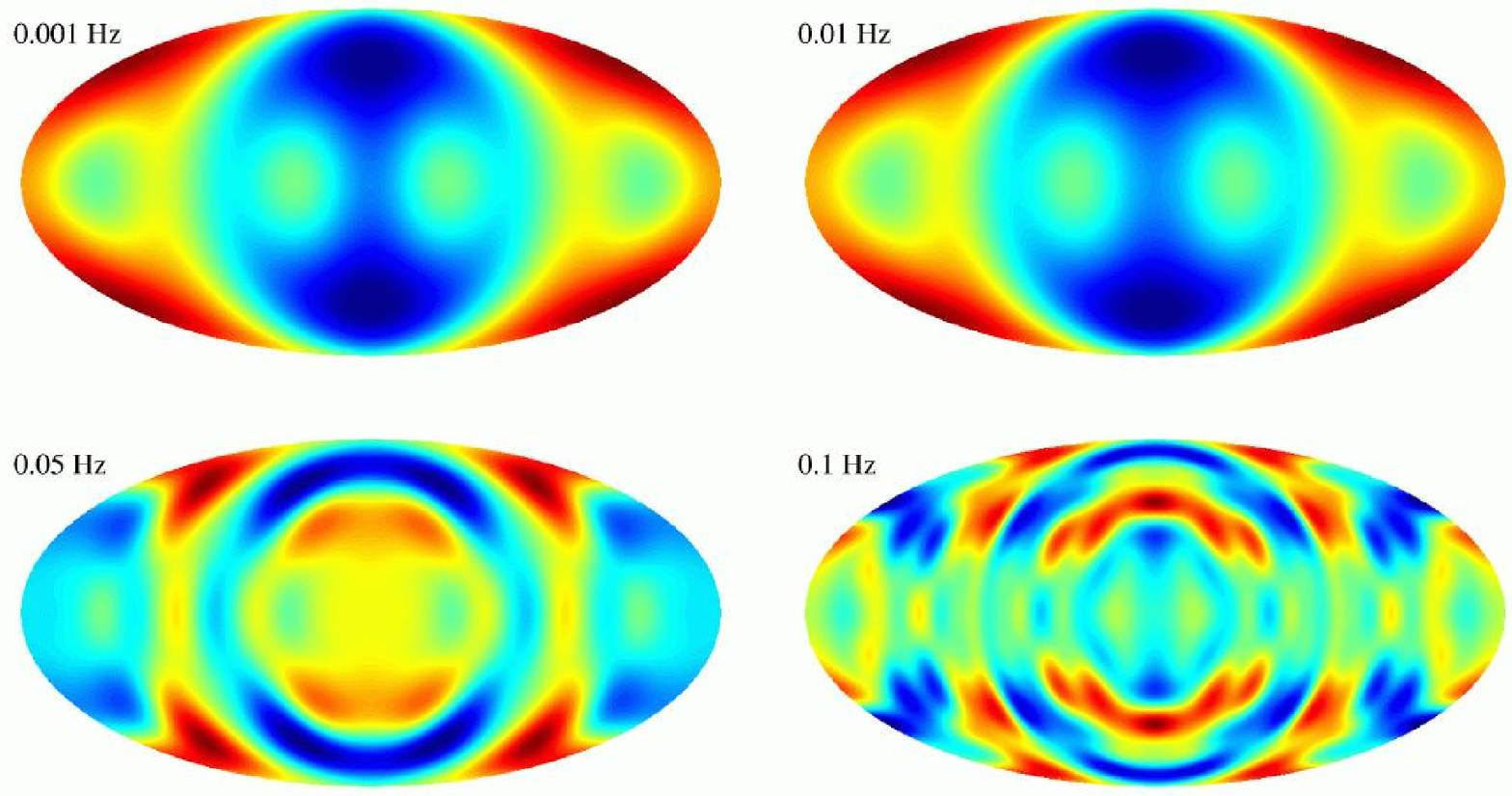}
\caption{The antenna pattern for two optimally cross-correlated LISA detectors in the detector
rest frame for a range of frequencies. Here $q_R=0$ and $q_I=1$.}
\end{figure} 

Combined with the rotation coefficients $\lambda_{k \ell m n}$, these give the
$\gamma_{k \ell m}$'s that we need to solve the deconvolution problem in the
low frequency limit. For example, the self-correlated LISA detector has
\begin{equation}
\hspace*{-0.1in} \begin{array}{lllll}
&\gamma_{000}=1  \quad 
&\gamma_{222}=\frac{3\sqrt{30}}{56} \quad
&\gamma_{121}=-\frac{3\sqrt{10}}{28} \quad
&\gamma_{020}=-\frac{\sqrt{5}}{28} \\ \bs
&\gamma_{-12-1} =\frac{3\sqrt{10}}{28}
&\gamma_{-22-2} =\frac{3\sqrt{30}}{56}
&\gamma_{044} = -\frac{27\sqrt{70}}{7168}
&\gamma_{444} = \frac{3\sqrt{70}}{3584}  \\ \bs
&\gamma_{844} = -\frac{\sqrt{70}}{21504}
&\gamma_{-14-3} = -\frac{9\sqrt{105}}{1792}
&\gamma_{343} = -\frac{\sqrt{105}}{896}
&\gamma_{743} = \frac{\sqrt{105}}{5376} \\ \bs
&\gamma_{-242} = \frac{9\sqrt{10}}{512}
&\gamma_{242} = \frac{3\sqrt{10}}{1792}
&\gamma_{642} = -\frac{\sqrt{10}}{512}
&\gamma_{-341} =  -\frac{3\sqrt{15}}{256} \\ \bs
&\gamma_{141} = \frac{5\sqrt{15}}{2688}
&\gamma_{541} = \frac{\sqrt{15}}{256}
&\gamma_{-440} = -\frac{15}{512}
&\gamma_{040} = -\frac{37}{5376}  \\ \bs
&\gamma_{440} = -\frac{15}{512}
&\gamma_{-54-1} = -\frac{\sqrt{15}}{256}
&\gamma_{-14-1} = -\frac{5\sqrt{15}}{2688}
&\gamma_{34-1} = \frac{3\sqrt{15}}{256}  \\ \bs
&\gamma_{-64-2} =-\frac{\sqrt{10}}{512}
&\gamma_{-24-2} =\frac{3\sqrt{10}}{1792}
&\gamma_{24-2} =-\frac{9\sqrt{10}}{512}
&\gamma_{-74-3} =  -\frac{\sqrt{105}}{5376} \\ \bs
& \gamma_{-34-3} = \frac{\sqrt{105}}{896}
& \gamma_{14-3} =\frac{9\sqrt{105}}{1792}
& \gamma_{-84-4} = -\frac{\sqrt{70}}{21504}
& \gamma_{-44-4 } = \frac{3\sqrt{70}}{3584}  \\ \bs
& \gamma_{04-4 } = -\frac{27\sqrt{70}}{7168} \, .
& & \\
\end{array}
\end{equation}

\begin{figure}[ht]
\vspace*{4.8in}
\includegraphics{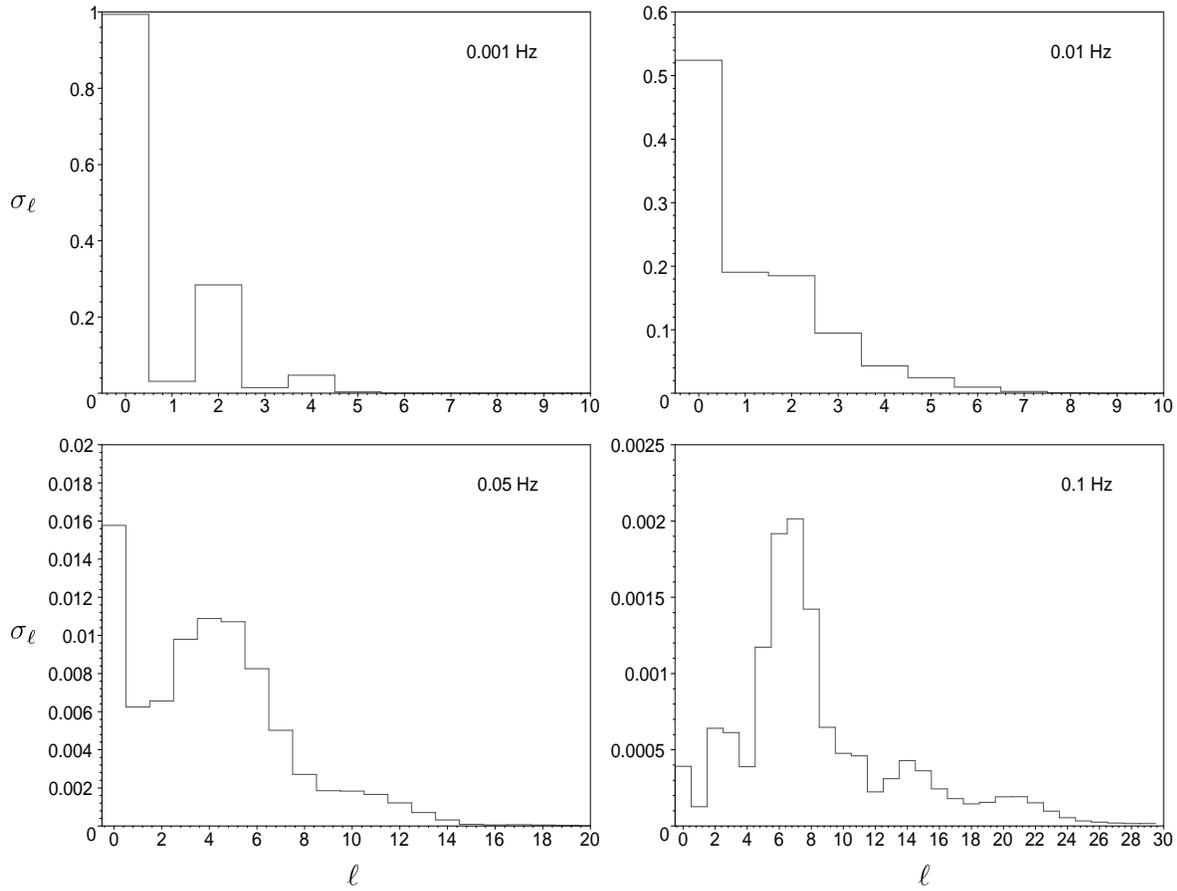}
\caption{The root-mean-square amplitudes of $\bar{a}_{\ell m}(f)$ for the
optimally cross-correlated LISA detectors at a range of
frequencies.}
\end{figure}

Thus, at very low frequencies, the inversion problem only involves $\vert k \vert \leq 8$
and multipoles with $\ell=0,2$ and $4$, which gives us 17 equations for 15 unknowns.
Writing out the convolution problem explicitly reveals that the system is simultaneously
under and over constrained. In other words, some individual multipoles will be very well
determined while others can only be given in combination with other multipoles. Once again
the degeneracy can be split by building additional detectors that make different antenna
sweeps. Alternatively, the LISA constellation could be re-position after several years of
data collection into a new orbit that is inclined with respect to the ecliptic.

At higher frequencies the LISA detectors are sensitive to more multipoles. Plots of the
root-mean-square amplitude, $\sigma_\ell(f)$, of the antenna multipoles $\bar{a}_{\ell m}$
are shown in Figure 6 for the optimally cross-correlated LISA detectors. By either
re-positioning the LISA constellation or by flying multiple constellations at different
inclinations, it will be possible to produce maps with a resolution approaching
$\ell=25$, which corresponds to an angular resolution of seven degrees.

\ack

I benefitted from discussion with Shane Larson, Ron Hellings and Bill Hiscock. I would
like to thank Kris Gorski for help with the {\it HEALPIX} software package.

\end{document}